# Giant longitudinal negative magneto-resistance under perpendicular magnetic field in $Bi_{2-x}Fe_xSe_{3-x}S_x$ Topological insulators

Rahul Singh[1], Vinod K. Gangwar[1], Abhishek Singh[1], A. K. Ghosh[2], Soma Banik[3], A. Lakhani[4], S. Patil[1] and Sandip Chatterjee[1,*]

[1]Department of Physics, Indian Institute of Technology (Banaras Hindu University), Varanasi-221005, India

[2]Department of Physics, Banaras Hindu University, Varanasi-221005, India

[3]Synchrotrons Utilization Section, Raja Ramanna Centre for Advanced Technology, Indore-452013, India.

[4]UGC-DAE Consortium for Scientific Research, Indore, Madhya Pradesh 452017, India

## Abstract

The magnetic, magneto-transport and ARPES studies of Fe and S co-doped $Bi_2Se_3$ were investigated. With doping concentration magneto-resistance (MR) gradually decreases and for a certain doping concentration giant negative MR is observed which persists up to room temperature. Magnetic measurement indicates that the negative MR is observed when ferromagnetism is induced with Fe doping. The magnetic ordering can be attributed with the RKKY interaction. Positive MR reappears with larger doping concentration which may be attributed to the decrease of FM ordering due to the turning off of the spin-orbit coupling leading to the destruction of non-trivial bulk state. This in-effect de-hybridizes the conduction band with the Fe spin. The ARPES data also indicates that above a critical doping concentration ($x>0.09$) the non-trivial bulk state is completely destroyed.

The magneto-transport properties in Topological insulators have attracted lot of attention due the interesting magnetoresistance behavior in these materials.[1,2] In fact, the topological surface states (TSS) exhibit many promising quantum phenomena associated with transport studies of topological insulators (TIs) *viz.* weak antilocalization (WAL)[3] and the two dimensional (2D) Subhnikov de-Haas (SdH) oscillation.[4] Moreover, positive non-saturating linear magnetoresistance (LMR) is observed in TIs.[5]

Recently, magneto transport study on $Bi_2Se_3$ epitaxial layer was executed at low temperature via angle dependent SdH oscillation[6] which results the features of both the bulk and surface states. Moreover, a strong anisotropic behavior was found in MR, depending on the orientation of current with respect to the magnetic field over a wide range of carrier concentration. A strong negative linear magnetoresistance (NLMR) was also observed while applying the magnetic field parallel to electric one.[7]

The recent surge of interest on Dirac Fermion in “3D TI” or “Weyl” semi metal[8] is based basically on their interesting topological properties that features the negative linear magnetoresistance in Weyl semi metal, when the magnetic and electric fields are coaligned. Quantum mechanical phenomena in which number of imbalance chiral fermions exist under an electric field[9,10] known as axial anomaly, is mainly responsible for NLMR. However, the large negative magnetoresistance (NMR) under perpendicular magnetic field in Topological insulator has not yet been reported. In a recent study[11] the NMR has been reported for S doped $Bi_2Se_3$ but the value is very low (<1%). Moreover, the low NMR has also been reported in $TlBi_{0.15}Sb_{0.85}Te_2$ system where at low temperature the resistivity is thermally activated.[12] However, origin of the NMR behavior either for parallel or perpendicular magnetic field is not yet perceived.

In this letter, we have reported the magnetoresistance behavior of $Bi_{2-x}Fe_xSe_{3-x}S_x$ system. It is observed that with increase of doping content, magnetoresistance gradually decrease and it reveals giant NMR at ~9% of doping. Nevertheless, with further increase of doping concentration, the positive MR reappears. This interesting finding confirms that the observation of NMR is not present only in Weyl semi metal but also exists in other Topological insulators.

The single crystal $Bi_{2-x}Fe_xSe_{3-x}S_x$ [with x =0.06, 0.09 and 0.12; we have denoted the doped samples as BiFeS-L, BiFeS-M and BiFeS-H respectively] were grown by melting a stoichiometric mixture with high purity Bi, Se, Fe and S elements which was sealed in evacuated ampoules. We took Fe and S in the equal stoichiometric ratio. The ampoule was heated up to 900˚C at 200˚C per hour and kept for 10 hours and then it was slowly cooled to 620˚C at 3˚C per hour. After that it was cooled down to 550˚C at 5˚C per hour and then it was naturally cooled down to room temperature. Thus obtained crystals were easily cleaved along (*00l*) direction. The electrical transport properties were carried out by using a Quantum Design physical properties measurement system (PPMS at 9T) and magnetic properties measurement were carried out by superconducting quantum interference device-vibrating sample magnetometer (SQUID-VSM) from quantum design.

The magneto-resistance (MR) as a function of a magnetic field at different temperatures of $Bi_{2-x}Fe_xSe_{3-x}S_x$ samples is shown in fig. 1. Magnetic field is applied along the perpendicular direction of the plane of the sample. We have defined MR as [{ρ(H)-ρ(0)}/ρ(0)]*100%. It is observed that MR value decreases with doping (x=0.06) and with further increase of doping content (x=0.09) large negative MR (NMR) throughout the whole temperature and magnetic field of measurements is observed. However, if the doping content is further increased positive MR reappears. In the inset we have shown the temperature variation of resistivity under different magnetic fields which supports the magnetic field variation of MR behavior.

Generally, the NMR in TIs is observed when applied magnetic field is co-aligned to the electric current and along the parallel direction to the surface.[13] In the present investigation the NMR is found when the magnetic field is perpendicular to the surface. Generally, the possible reasons of NMR are: Kondo effect quenching,[14-15] existence of ferromagnetic metallic state,[16] and existence of chiral anamoly.[17-18] The chiral anamoly is observed only when magnetic field is parallel to the electric current. However, in the present investigation the magnetic field is perpendicular to the surface and hence perpendicular to the electric current. Therefore, chiral anomaly is not applicable for the present case. Furthermore, no signature of Kondo effect is observed for the BiFeS-M sample both from the transport and magnetic measurements.

So far, most of the NMR effects found in 3D TIs are due to the coexistence of weak localization (WL) and weak anti localization effects under a low magnetic field.[1] However, in the present investigation for the BiFeS-M sample, NMR cannot be due to WL effect as the WL induced NMR saturates at a magnetic field ~1 T because of the smaller magnetic length over the phase coherence length in these Topological insulators.[19,20] However, in the present investigation MR is not saturated even at 8T magnetic field. Moreover, the NMR persists until 300K, too high temperature to exist WL. Furthermore, in a recent paper[21] it has been proposed that the observed NMR might be due to the Zeeman splitting which originates due to the spin-orbit coupling on the surface and in the parallel direction to the applied magnetic field. But in the present case the NMR is observed when the magnetic field is perpendicular to the electric field. Therefore, Zeeman splitting is not the origin of NMR in our case.

In a recent paper[12] Breunig et al. have shown in $TlBi_{0.15}Sb_{0.85}Te_2$ Topological insulator the NMR under perpendicular magnetic field is due to the electron puddles. But for electron or hole puddles to occur there should be thermally activated conduction. But in our case all the samples (undoped and S doped) show completely metallic behavior. Therefore, most likely, electron or hole puddles are not the origin of negative MR in present case. The most interesting observation in this system is the reappearance of positive MR with further increase of doping concentration (BiFeS-H).

In order to further investigate the origin of the observed MR behavior, we have also measured the magnetization with magnetic field at different temperatures (Fig. 4). It is observed that undoped sample shows diamagnetic behavior but as the doping content increases it shows the ferromagnetic ordering. For BiFeS-M co-doped sample maximum magnetization is observed. As the doping content increases further (BiFeS-H) the magnetization value is decreased. The ferromagnetic behavior also observed even at 300K for the BiFeS-M sample. Therefore, it may be concluded that the negative MR is observed because of the existence of ferromagnetic ordering. In this case the origin of the magnetic ordering might be the RKKY interaction as has been suggested by Kim et al[22] from theoretical calculation. In fact, in the present case the S doping already shifts the Fermi surface to the conduction band [11] and hence the Fe d orbital hybridizes with the conduction band. Initially, for low Fe doping the distance between Fe ions are very large and because of the RKKY interaction it shows weak FM ordering.[22] With further increase of Fe content the FM ordering further increases because of the change in $k_F$ (due to S doping) and also of the change in Fe-Fe distances. But for the highest doped sample, the S already destroys the non-trivial surface state[11] by turning off the spin-orbit coupling.[23] Therefore, there will be no more any band inversion and conduction band without any band inversion cannot work as mediating channels for magnetic interaction. As a matter of fact, Fe atoms will have independent paramagnetic states without any preferential spin direction and hence, ferromagnetism is decreased.

In order to further investigate we have also measured the ARPES for all the samples. In fig. 3 we have shown the ARPES spectra collected for BiFeS-L, BiFeS-M and BiFeS-H. All these spectra are normalized and then plotted using the same color scale for intensity comparison. For BiFeS-L we observe the bulk conduction band (BCB), the surface state (SS) and the bulk valence band (BVB) for the compound. Such a SS has been observed in $Bi_2Se_3$ previously and is known to possess the topological protection which arises from the topology of the bulk electronic structure of $Bi_2Se_3$.[23] These SS possess a linear dispersion (with a Dirac cone structure) as that expected for a massless particle and therefore it comprises of the massless Dirac fermions as opposed to normal fermions.[23] Due to the topological protection the SS is immune to scattering due to non-magnetic disorders introduced on the surface of $Bi_2Se_3$ whereas the introduction of magnetic disorder on the surface results in the loss of the topological protection of the SS due to the breaking of the time reversal symmetry (TRS) thereby opening a gap at the Dirac point.[24] In this study, however, we are attempting to introduce magnetic dopants (i.e. Fe atoms) within the bulk of the $Bi_2Se_3$ matrix and intending to study its effect on the evolution of the SS. So in this case of bulk doping the topology of the bulk electronic structure would be modified and hence it is curious to study its effect on the SS. All the collected ARPES spectra show a portion of the BCB within the occupied region of the ARPES image due to the *n*-type doping of the $Bi_2Se_3$ matrix arising out of the intrinsic Se vacancies within it. Furthermore the Fe and S co-doping leads to the additional *n*-type doping as is evident from the gradual shift of the bulk bands towards higher binding energies (BE) with doping. The absence of a band gap in the electronic structure of Dirac fermions on the surface of BiFeS-L shows that the doped magnetic Fe atoms in the sample are not yet sufficient to affect the Dirac cone dispersion at the sample surface. In figs. (d) and (e) we show the second derivative maps of the ARPES intensity (for BiFeS-L and BiFeS-H respectively) which emphasize the electronic bands within an ARPES image. Accordingly, the Dirac cone structure of the Dirac fermions is emphasized in BiFeS-L; however for BiFeS-H the Dirac cone is not visible. Instead, the contours that we observe in case of BiFeS-H arise due to the peripheries of the high intensity regions of its ARPES spectra which in fact correspond to its bulk electronic structure. There is no trace of the Dirac cone in $Bi_2Se_3$-H as is the case in BiFeS-L. The loss of intensity close to the Dirac point is clearly evident from fig. 3(f) where we show the energy distribution curves (EDC) at the Γ-point of the Brillouin zone extracted from figs 3(a), (b) and (c). A comparison between them clearly shows that the ARPES intensity around the region close to the Dirac point (inside the dashed rectangle) reduces as we increase the doping. In BiFeS-H the intensity is negligible and hence there the gap has been fully opened up. We thus conclude the absence of SS and thereby the Dirac fermions in the case of high doping. This is not at all surprising since we are doping magnetic atoms not just at the surface of $Bi_2Se_3$ but right in the bulk in which case we are tampering with the topology of its bulk electronic structure upon which the very existence of the topologically protected SS rests. Thus with the increase in the doping concentration above that of BiFeS-L we observe the absence of the Dirac dispersion and hence we conclude that the non-trivial topology of the bulk electronic structure of $Bi_2Se_3$ matrix has been destroyed. This corroborates well with the SdH oscillations discussed below where the contribution to those oscillations arising from the Dirac fermions is absent. Thus from ARPES studies we conclude that the bulk contribution to the electronic properties of the compound increases with doping which is also consistent with the magnetotransport results.

In order to further support the ARPES results we have also shown the fast Fourier transforms (FFTs) of the quantum oscillation (Fig.4). For undoped and low doped we can distinguish the bulk state (lowest frequency peak) and surface states (higher frequency peaks). For the higher doped samples also several peaks are observed. Among these the peaks of low width and low intensity might be background or might be due to the contribution of channels of lower mobility. For BiFeS-H, it is observed the prominent peaks at higher frequencies are harmonics of the peak at the lowest frequency. Therefore, it is clear that in heavily doped samples bulk contribution is dominant. The Landau-level fan diagram (Landau index vs. 1/B, B being the magnetic field) of all the samples (shown in Fig. 6) shows an intercept at ∼0.47 for the undoped and ~0.41 for the lowest doped samples which indicates that the Dirac fermions (with additional Berry phase $\pi$) dominate the transport properties. It is found that as the doping content increases the deviation of the intercept from 0.5 also increases (and becomes close to zero) revealing that in the transport properties the contribution of Dirac fermion decreases, while the contribution of normal fermion increases. This also clearly indicates that bulk conduction gradually dominates over surface conduction with doping and finally for BiFeS-H sample no signature of non-trivial surface state is observed. Therefore, the magneto-transport property is completely consistent with the ARPES study.

It is clear from both the ARPES and magneto-transport studies that with increase of doping content the Fermi energy moves into the conduction band. In fact, this happens because of the S doping, as each S by replacing Se donates two electrons. This we already have shown in our previous study.[11] It is observed that increase of S content gradually decreases the MR and at some doping concentration it becomes negative.[11] But value of the negative MR is very low (<1%). Moreover, it has already been reported that Fe doping in TI induces FM ordering.[24-26] But the ordering temperature is very low (<25K). In the present investigation the FM ordering is observed even at 300K. Previously some controversial results were reported about gap opening of the surface Dirac band with Fe doping in Topological Insulator. In some paper[24] gap opening has been reported whereas in some reports from ARPES studies[26,27] it has been claimed that no gap is opened with magnetic ion doping. In fact, the null results are consistent with the report by Kim et. al.[26] where it has been suggested that the positions of the Fermi level moves far from the Dirac point when Fe is doped inducing ferromagnetism without opening the gap. As a matter of fact, the bulk doping of magnetic ions is more effective than the surface doping for controlling the topological characters. As has been shown in the present investigation from magnetic, magneto-transport and ARPES studies that for low Fe doping the band gap is not opened but FM is induced clearly indicating the bulk doping. However, the gapped surface state can be realized by bulk doping of magnetic ions.[7] We observe in our case also when doping content is increased surface gap is opened. But even at 9% doping the existence of Dirac cone is observed from the ARPES. It is obvious that with increase of FM ordering the negative MR enhances and finally for highest doped sample positive MR reappears because of the decrease of ferromagnetism as has been discussed above. Similar positive linear MR is observed in non-Topological $Bi_{1-x}Sb_x$ material.[28] It is also obvious from above discussion that the large FM ordering is the simultaneous effect of both Fe and S and this FM ordering is the origin of negative MR. NMR persists upto room temperature as FM ordering exists even at RT although the magnetization value decreases as compared to low temperature (due to thermal agitation).

**Conclusion**:

The magnetic, magneto-transport and ARPES studies of $Bi_{2-x}Fe_xSe_{3-x}S_x$ were carried out. It is observed that with increase of doping concentration MR gradually decreases and for x=0.09 it shows giant negative magnetoresistance even upto room temperature. The negative MR is observed when the system is in the FM state. The origin of FM ordering is attributed to the RKKY interaction. At large doping (>9%) the non-trivial surface state is completely destroyed which in effect destroys the spin-orbit coupling. As a matter fact, conduction band cannot hybridized with the Fe d orbitals as the conduction band is no more inverted and the Fe d orbitals will not have any particular orientation. This is reason of the reappearance of positive MR with larger doping. The ARPES data indicates that above a critical doping concentration (x>0.09) the non-trivial bulk state is completely destroyed.

**Acknowledgement**: The authors are grateful to Dr. Tapas Ganguly, RRCAT, Indore, for providing facility for ARPES measurement. The authors are also grateful to CSR-DAE, Indore , and CIFC, IIT(BHU) for providing facilities for magnetotransport measurements and magnetic measurement respectively.

Figure Captions:

Fig.1: Magnetic field variation of Magneto-resistance at different temperatures of $Bi_2Se_3$, BiFeS-L, BiFeS-M and BiFeS-H. Inset: Temperature variation of resistivity at different magnetic fields.

Fig.2: Magnetic field variation of magnetization of $Bi_2Se_3$, BiFeS-L, BiFeS-M and BiFeS-H at different temperatures.

Fig. 3: (a), (b) and (c) shows the ARPES spectra of BiFeS-L, BiFeS-M and BiFeS-H respectively collected by photons of energy 21.2 eV. (d) and (e) shows the second derivative of ARPES images for BiFeS-L and BiFeS-H respectively. The green dots in (d) depicts the Dirac cone structure. (f) shows the comparison of the EDC's at the Γ-point of the Brillouin zone for all the samples.

**Fig.4:** Landau level index (obtained from MR quantum oscillations) as a function of inverse magnetic field of $Bi_2Se_3$, BiFeS-L, BiFeS-M and BiFeS-H. Inset (right): Quantum oscillation obtained from magnetic field dependence MR data. Inset (left): First Fourier transform of quantum oscillations.

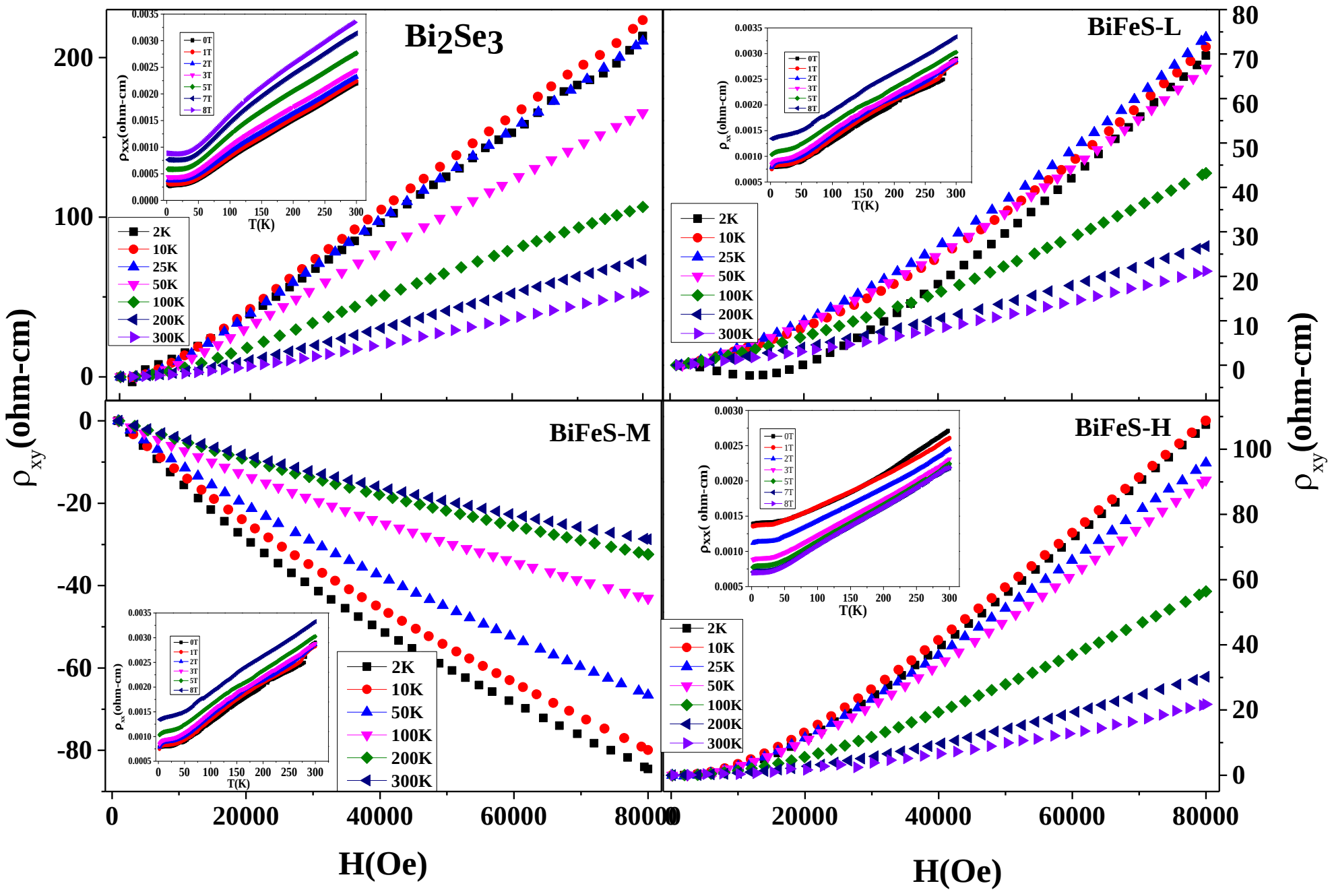

Fig.1

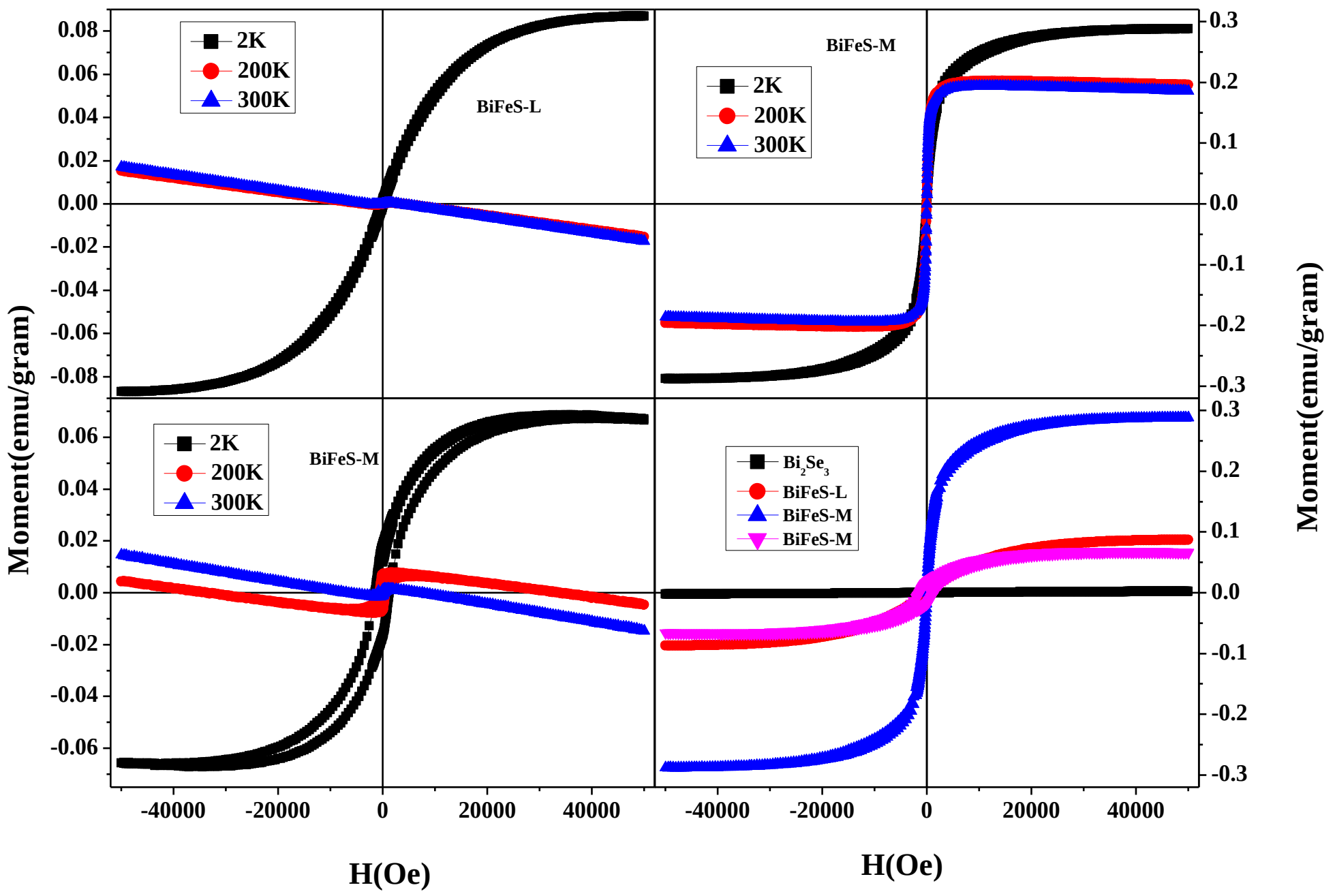

Fig.2

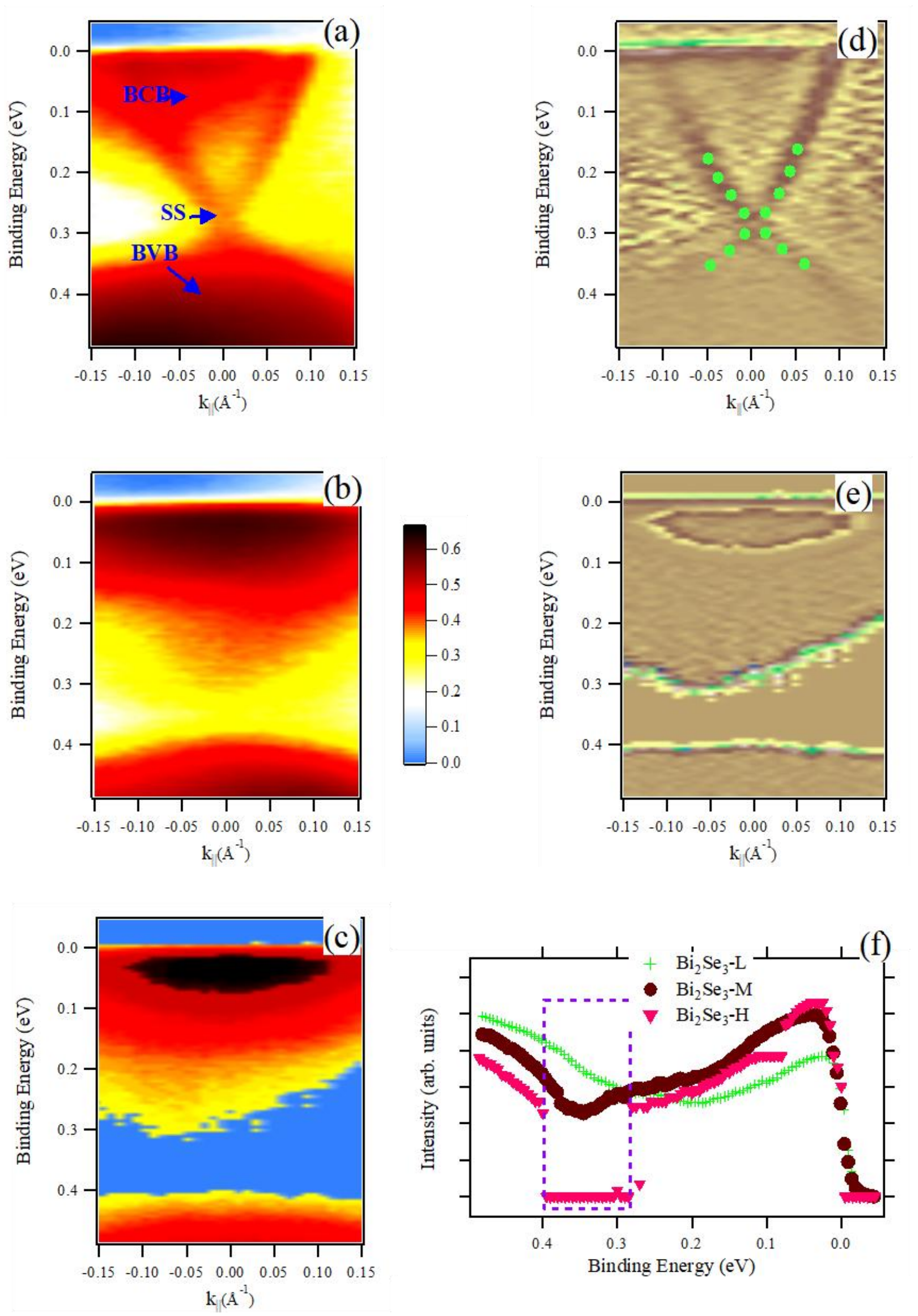

**Fig.3**

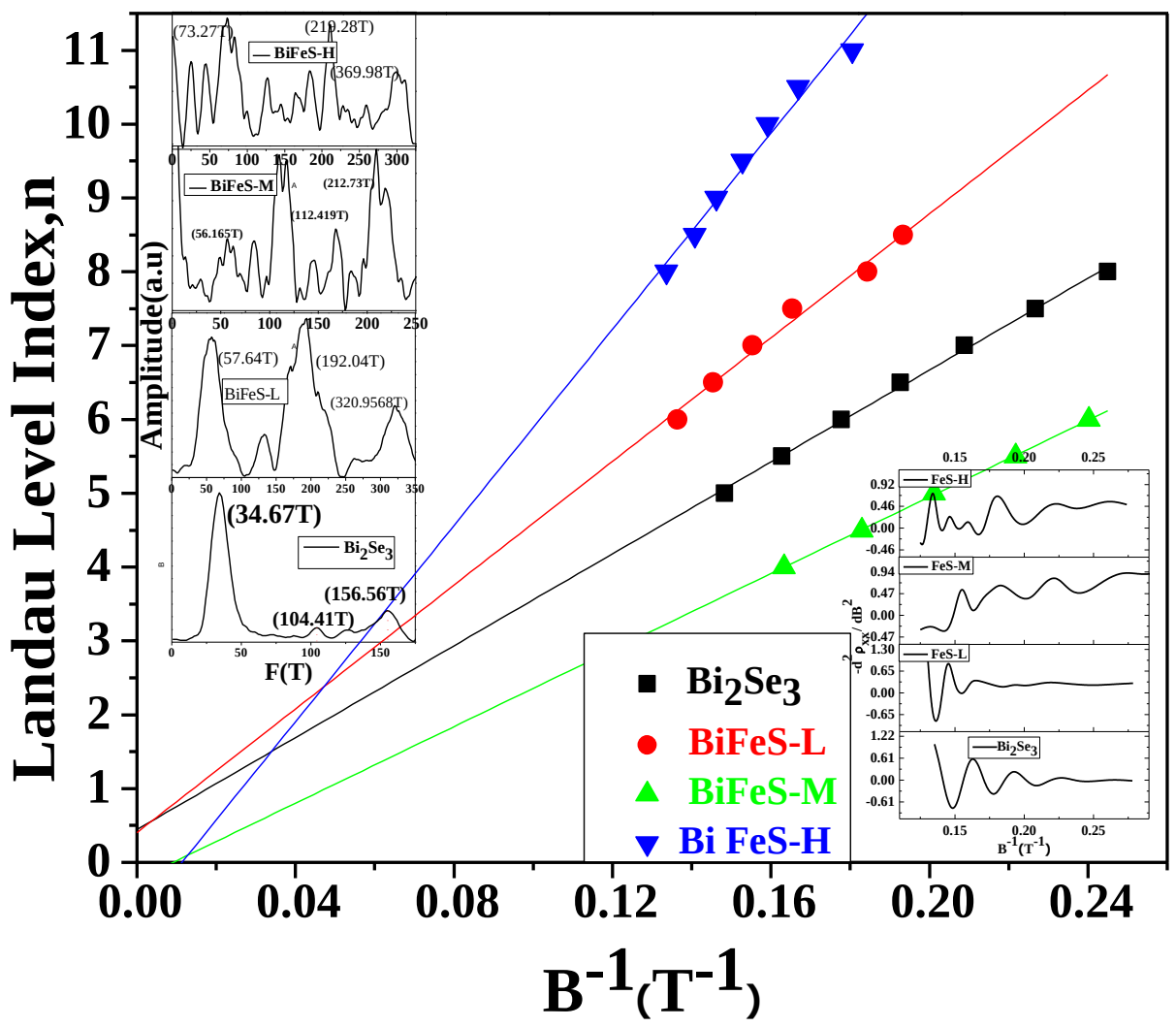

**Fig.4**